\documentstyle[11pt,graphicx]{article}
\input epsf 

\begin{document} 

%
\newcommand\rd{{\rm d}}
\newcommand\p{\partial}
\newcommand\nab{\nabla}
\newcommand\pr{\prime}
\newcommand\q{\quad}
\newcommand\qq{\qquad}
\newcommand\r{\right}
\renewcommand\l{\left}

\newcommand\e{{\epsilon}}
\newcommand\ga{{\gamma}}
\newcommand\th{{\vartheta}}

\newcommand\bA{{\bf A}}
\newcommand\bB{{\bf B}}
\newcommand\bF{{\bf F}}
\newcommand\bJ{{\bf J}}
\newcommand\bX{{\bf X}}
\newcommand\bu{{\bf u}}
\newcommand\bx{{\bf x}}
\newcommand\bnu{{\mbox{\boldmath $\hat\nu$}}}
\newcommand\bom{{\mbox{\boldmath $\omega$}}}
\newcommand\bxi{{\mbox{\boldmath $\xi$}}}
\newcommand\Mu{{M}}
\newcommand\Nu{{N}}
\newcommand\bR{{\bf R}}
\newcommand\ba{{\bf a}}
\newcommand\bb{{\hat{\bf b}}}
\newcommand\bn{{\hat{\bf n}}}
\newcommand\bt{{\hat{\bf t}}}

\newcommand\cC{{\cal C}}
\newcommand\cD{{\cal D}}
\newcommand\cF{{\cal F}}
\newcommand\cN{{\cal N}}
\newcommand\cR{{\cal R}}
\newcommand\cS{{\cal S}}
\newcommand\cT{{\cal T}}
\newcommand\cU{{\cal U}}

\newcommand\beo{{{\hat{\bf e}}_1}}
\newcommand\bet{{{\hat{\bf e}}_2}}
\newcommand\ber{{{\hat{\bf e}}_r}}
\newcommand\beth{{{\hat{\bf e}}_\theta}}

\newcommand\bethr{{{\hat{\bf e}}_{\vartheta_{R}}}}
\newcommand\fc{{f_c}}
\newcommand\ft{{f_\tau}}
\newcommand\s{{\hat{s}}}
\newcommand\A{{\tilde{A}}}
\newcommand\B{{\tilde{B}}}
\newcommand\C{{\tilde{C}}}
\newcommand\D{{\tilde{D}}}
\newcommand\E{{\tilde{E}}}
\newcommand\F{{\tilde{F}}}
\newcommand\G{{\tilde{G}}}
\renewcommand\H{{\tilde{H}}}
\newcommand\I{{\tilde{I}}}
\newcommand\J{{\tilde{J}}}
\newcommand\Bs{{B_s}}
\newcommand\bBa{{{\bf B}_a}}
\newcommand\bBm{{{\bf B}_m}}
\newcommand\Bth{{B_\theta}}
\newcommand\thr{{\vartheta_R}}
\newcommand\bTn{{{\hat{\bf T}}_2}}
\newcommand\bTb{{{\hat{\bf T}}_3}}

\newcommand\NN{{\hbox{I\kern-.14em{N}}}}
\newcommand\RR{{\hbox{I\kern-.14em{R}}}}
\newcommand\ZZ{{\hbox{I\kern-.14em{Z}}}}

\newtheorem{defini}{Definition}[section]
\newtheorem{propo}{Proposition}[section]
\newtheorem{teo}{Theorem}[section]
\newtheorem{coro}{Corollary}[section]
\newtheorem{osse}{Remark}[section]
\newtheorem{nota}{Notation}[section]
\newtheorem{notaz}{Notations}[section]
\newtheorem{ese}{Exercise}[section]
\newtheorem{comme}{Comment}[section]
\newtheorem{esempio}{Example}[section]
\newtheorem{proble}{Problem}[section]

\newcommand{\qed}{{\par\hfill $\Box$\medskip}}


\pagestyle{empty}
\parindent=0mm
\null\vspace{-20mm}
\begin{center}
{\LARGE\bf Kinetic energy of vortex knots and unknots}\\ 
\vspace{10mm}
\large{\textsc{Francesca Maggioni}${\;}^{1,}$\footnote{Corresponding author: \tt francesca.maggioni@unibg.it}}, 
\textsc{S.Z.~Alamri}${\;}^{2,3}$
\textsc{C.F.~Barenghi}${\;}^{3}$
\textsc{Renzo L. Ricca}${\;}^{4}$\\

\vspace{3mm}
{\it ${}^{1}$ Dept. Mathematics, Statistics, 
Computer Science \& Applications, U. Bergamo}\\
{\it Via dei Caniana 2, 24127 Bergamo, ITALY}\\
\vspace{1mm}
{\it ${}^{2}$ Department of Applied Mathematics, 
             College of Applied Science}\\
{\it P.O. Box 344, 
	   Taibah University, Al-Madinah Al-Munawarah, SAUDI ARABIA.}\\
\vspace{1mm}
	   {\it ${}^{3}$ School of Mathematics and Statistics
             Newcastle University}\\
{\it Newcastle upon Tyne, 
	     NE1 7RU, U.K.}\\
and\\
{\it ${}^{4}$ Dept. Mathematics \& Applications, U. Milano-Bicocca}\\
{\it Via Cozzi 53, 20125 Milano, ITALY}\\ 
\vspace{5mm}
\vspace{10mm}
ABSTRACT\\
\end{center}
New results on the kinetic energy of ideal vortex filaments in the
shape of torus knots and unknots are presented.  These knots are given
by small-amplitude torus knot solutions (Ricca, 1993) to the Localized
Induction Approximation (LIA) law.  The kinetic energy of different
knot and unknot types is calculated and presented for comparison.
These results provide new information on relationships between
geometry, topology and dynamics of complex vortex systems and help to
establish possible connections between aspects of structural
complexity of dynamical systems and vortical flows.
\vfill
%
%

\section{Vortex motion under Biot-Savart and LIA law}
The present work represents a natural extension of previous work
(Ricca {\em et al.}, 1999) on vortex torus knots and unknots.  In this
paper we present new results on the kinetic energy of these vortex
systems and investigate possible relationships between energy and
complexity of such structures.

We consider vortex motion in an ideal, incompressible (constant density)
fluid in an unbounded domain $\cD\subseteq\RR^{3}$. 
The velocity field $\bu=\bu(\bx,t)$, smooth 
function of the vector position $\bx$ and time $t$, satisfies
\begin{equation}
    \nabla\cdot\bu=0\quad \textrm{in } \cD\ ,\qquad \bu\to 0\quad 
    \textrm{as}\ \bx\rightarrow\infty\ ,
    \label{bu}
\end{equation}
and the vorticity $\bom$ is defined by
\begin{equation}
    \bom=\nabla\times\bu\ ,\qquad 
    \nabla\cdot\bom=0\quad \textrm{in}\ \cD\ .
    \label{bom}
\end{equation}
In absence of viscosity fluid evolution is governed by the Euler's
equations and vortical flows obey Helmholtz's conservation laws
(Saffman, 1992). Transport of vorticity is govern by 
\begin{equation}
    \frac{\partial\bom}{\partial t}
    =\nabla\times\left(\bu\times\bom\right)\ ,
    \label{vortrans}
\end{equation}
whose formal solutions are given by the Cauchy equations
\begin{equation}
    \omega_i(\bX,t)=
    \omega_j(\ba,0)\frac{\partial X_i}{\partial{\rm a}_j}\ ,
    \label{cauchy}
\end{equation}
that encapsulate both convection of vorticity from the initial 
position $\ba$ to $\bX$, and the simultaneous rotation and distortion 
of the vortex elements by the deformation tensor $\p X_i/\p{\rm a}_j$.  
Since this tensor is associated with a continuous deformation of the
vortex elements, vorticity is mapped continuously from the initial 
configuration $\bom(\ba,0)$ to the final state $\bom(\bX,t)$; hence,
Cauchy equations establish a topological equivalence between initial 
and final configuration by preserving the field topology. In 
absence of dissipation, physical properties such as kinetic energy, 
helicity and momenta are conserved along with topological quantities 
such as knot type, minimum crossing number and self-linking number 
(Ricca \& Berger, 1996).

The kinetic energy per unit density $T$ is given by  
\begin{equation}
    T=\frac{1}{2}\int_{V}\|\bu\|^2\,\rd\bx^{3}=cst.\ ,
    \label{kinetic}
\end{equation}
where $V=V(\cD)$ is the volume of the ambient space $\cD$.  We assume
that we have only one vortex filament $\cF$ in isolation, where $\cF$
is centred on the axis $\cC$ of equation $\bX=\bX(s)$ ($s$ arc-length
on $\cC$).  The filament axis is given by a smooth (at least $C^{2}$),
simple (i.e. without self-intersections), knotted space curve.  The
filament volume is given by $V(\cF)=\pi a^{2}L$, where $L=L(\cC)$ is
the axis length and $a$ is the radius of the vortex core, which is
assumed to be uniformly circular all along $\cC$.  Vortex motion is
governed by the Biot-Savart law (BS for short) given by
\begin{equation}
  \bu(\bx)=\frac{\Gamma}{4\pi}\oint_{\cC}
  \frac{\bt\times(\bx-\bX(s))}{\|\bx-\bX(s)\|^{3}}\,\rd s\ ,
  \label{bs}
\end{equation}
where vorticity ($\bom=\omega_{0}\bt$) is expressed through the 
circulation $\Gamma$, $\omega_{0}$ being a constant and $\bt=\bt(s)$ 
the unit tangent to $\cC$.  Since the
Biot-Savart integral is a global functional of the vortex
configuration, that takes into account of the induction effects of
every element of the vortex, analytical solutions in closed form,
other than the classical solutions associated with rectilinear,
circular and helical geometry, are very difficult to obtain.
Considerable analytical progress, however, has been done for the
Localized Induction Approximation (LIA for short) law.  This equation,
first derived by Da Rios (1905) and independently re-discovered by
Arms \& Hama (1965) (see the review by Ricca, 1996), is obtained by a
Taylor's expansion of the Biot-Savart integrand about a point on the
vortex filament axis $\cC$ (see, for instance, Batchelor, 1967, \S 
7.1). By neglecting the rotational component of the self-induced 
velocity (that does not contribute to the displacement of the vortex 
in the fluid) and non-local terms, the LIA equation takes the 
simplified form 
\begin{equation}
  \bu_{\rm LIA}=\frac{\Gamma c}{4\pi}\ln{\delta}\,\bb\propto c\bb\ ,
  \label{lia}
\end{equation}
where $c=c(s)$ is the local curvature of $\cC$, $\delta$ is a measure 
of the aspect ratio of the vortex and $\bb=\bb(s)$ is the unit 
binormal vector at the point $\bX=\bX(s)$ of $\cC$. 

\section{LIA torus knots under Biot-Savart evolution}
\begin{figure}[t]
\centering
\includegraphics[width=0.7\textwidth]{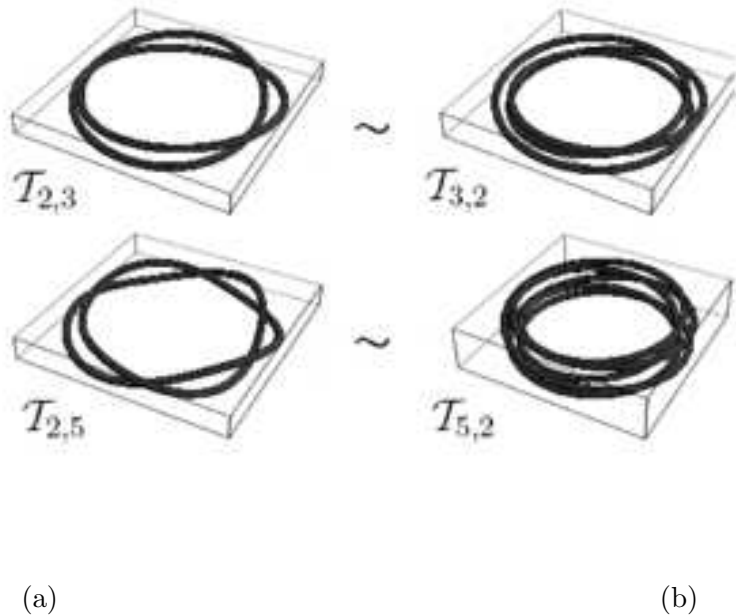}
\vskip3truemm\noindent
\centerline{\hspace{35mm}(a)\hfill(b)\hspace{35mm}}
\caption{Examples of torus knots $\cT_{p,q}$ given by solution eqs. 
(\ref{sols}) for (a) winding number $w>1$, and (b) winding number $w<1$.
Knots on the same row have different geometry, but both represent 
the same knot type, hence they are topologically equivalent: 
$\cT_{2,3}\sim\cT_{3,2}$ and $\cT_{2,5}\sim\cT_{5,2}$.}
\label{knots}
\end{figure}

\begin{figure}[t]
\centering
\includegraphics[width=0.7\textwidth]{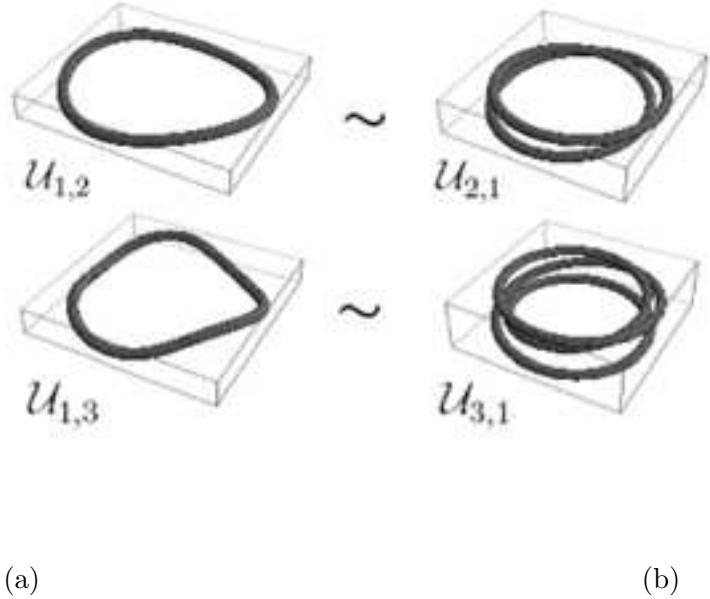}
\vskip3truemm\noindent
\centerline{\hspace{35mm}(a)\hfill(b)\hspace{35mm}}
\caption{Torus unknots given by solution eqs. (\ref{sols})
for (a) winding number $w>1$, and (b) winding number $w<1$.
All these unknots have different geometry, but they are all 
homeomorphic to the standard circle $\cU_0$.}
\label{unknots}
\end{figure}

We consider a particular family of vortex configurations in the shape
of torus knots in $\RR^{3}$.  These are given when $\cC$ is a torus
knot $\cT_{p,q}$ ($\{p,q\}$ co-prime integers), i.e. a closed curve
embedded on a mathematical torus, that wraps the torus $p$ times in
the longitudinal (toroidal) direction and $q$ times in the meridian
(poloidal) direction (see Figure \ref{knots}).  When $\{p,q\}$ are one
a multiple of the other the curve is no longer knotted and forms an
unknot (homeomorphic to the circle $\cU_\circ$) that, depending on the
value of $p$ and $q$, takes the shape of a curve wound by $m$ turns
helically around the torus either longitudinally, forming a toroidal
coil $\cU_{m,1}$, or meridionally, hence forming a poloidal coil
$\cU_{1,m}$ (see Figure \ref{unknots}).  When $\{p,q\}$ are not even
integers, but rational, $\cT_{p,q}$ is no longer a closed knot, but
the curve may fill the toroidal surface completely.  The ratio $w=q/p$
denotes the \emph{winding number} and $Lk=pq$ the \emph{self-linking
number}, two topological invariants of $\cT_{p,q}$.  Note that for
given $p$ and $q$ the knot $\cT_{p,q}$ is topologically equivalent to
$\cT_{q,p}$, that is $\cT_{p,q}\sim\cT_{q,p}$ (i.e. they are the same
knot type), but their geometry (by standard embedding on the torus) 
is different.

Now, let us identify the vortex filament axis with $\cT_{p,q}$ (more 
loosely, we shall refer to $\cT_{p,q}$ to denote the torus knotted 
vortex filament), and consider this vortex evolution in the ambient 
space. Geometric information affects dynamics, and hence energy, 
through the BS law (\ref{bs}), or the LIA law (\ref{lia}), and we
want to investigate how different geometries of same knot types affect
dynamical properties such as kinetic energy.

Existence of torus knot solutions to LIA were found by Kida 
(1981), who provided solutions in terms of elliptic integrals. 
By re-writing LIA in cylindrical polar coordinates 
$(r,\alpha, z)$ and by using standard linear perturbation techniques
Ricca (1993) found small-amplitude torus knot solutions (asymptotically 
equivalent to Kida's solutions) expressed in closed form. The solution
curve, written in parametric form in terms of the arc-length $s$, are
given by
\begin{equation}
\left\{ 
\begin{array}{lcl}
  r= r_0 +\epsilon\sin(w\phi)\ ,\\[2mm]
  \alpha= \displaystyle{\frac{s}{r_0} +\frac{\epsilon}{wr_0}
    \cos(w\phi)}\ , \\[2mm]
  z=\displaystyle{\frac{t}{r_0}
  +\epsilon\left(1+\frac{1}{w^2}\right)^{1/2}
    \cos(w\phi)}\ ,
\end{array} 
\right.
\label{sols}
\end{equation}
where $r_0$ is the radius of the torus circular axis and
$\epsilon\ll1$ is the inverse of the aspect ratio of the vortex.
Since LIA is strictly related to the Non-Linear Schr\"{o}dinger (NLS)
(see Ricca, 1996 for details), torus knot solutions (\ref{sols})
correspond to helical travelling waves propagating along the filament
axis, with wave speed $\kappa$ and phase $\phi=(s-\kappa t)/r_0$.  As
a result, the vortex moves in the fluid as a rigid body, with a
propagation velocity along the torus central axis (given by the
$\dot{z}$-component, associated with the $z$-component of eqs.
\ref{sols}) and a uniform, helical motion (given by the radial and
angular velocity component, associated with the corresponding
components of \ref{sols}) of the knot strands along, and around, the
torus circular axis.  In physical terms, these waves provide an
efficient mechanism for the transport of kinetic energy and momenta
(and infinitely many other conserved quantities associated with NLS)
in the bulk of the fluid.

By using eqs.  (\ref{sols}) we have the following linear stability
result.

\begin{teo} (Ricca, 2005). Let $\cT_{p,q}$ be a small-amplitude vortex
torus knot under LIA. Then $\cT_{p,q}$ is steady and stable under
linear perturbations iff $q>p$ ($w>1$).
\end{teo}

This result provides a criterium for LIA stability of vortex knots,
and it can be easily extended to hold true for torus unknots (i.e.
toroidal and poloidal coils).  This has been confirmed to hold true
for several knot types tested by numerical experiments (Ricca \emph{et
al.}, 1999).  More interestingly, LIA unstable torus knot solutions
were found to be stabilized by the global induction effects of the BS
law. This is an unexpected result that has motivated further work and 
current research, which is in progress.


\section{Length versus complexity of LIA torus knots}
It is interesting to compare the total length of torus knot solutions
given by eqs. (\ref{sols}). Note that under LIA the vortex moves in 
the binormal direction, that is everywhere orthogonally to the unit 
tangent to $\cC$; the total length of the vortex is therefore 
conserved under LIA. This is clearly one limitation of the model, 
since it is well-known (Batchelor, 1967; Saffman, 1992), that in 
three-dimensional flows vortices do stretch --- a property 
measured by the increase of \emph{enstrophy} $\Omega$, that 
under LIA reduces to
\begin{equation}
    \Omega=\int_{V}|\bom|^{2}\,\rd\bx^{3}
          =\Gamma\omega_{0}\oint_{\cC}\rd s
	  =\Gamma\omega_{0}L=cst.\ .
    \label{enstrophy}
\end{equation}
Numerical work has been carried out
by setting $r_0=1~\rm cm$ (our code works with CGS units),
$\epsilon=0.1$, $\Gamma=10^{-3}~\rm cm^2/s$, (the value expected for
superfluid $^4$He), $\delta\gg1$ (in our code 
$\delta=2\cdot 10^{8}/e^{1/2}$) and, to study unknots, by replacing
$(1-1/w^2)$ with $|1-1/w^2|$ in eqs.  (\ref{sols}). A reference 
vortex ring $\cU_{0}$ is taken with radius $r_0=1~\rm cm$. 
Convergence has been tested in space and time by modifying 
discretization points and time steps. 

\begin{figure}[t!] 
\centering    
\includegraphics[width=0.7\textwidth]{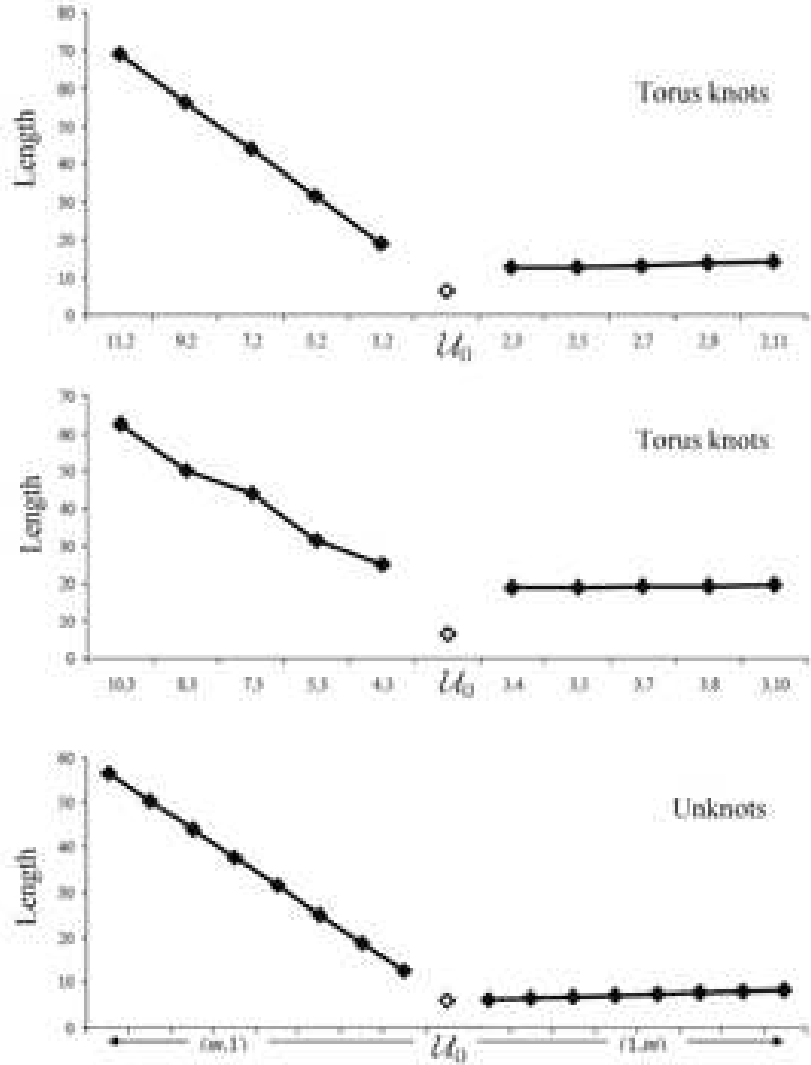}  
\caption{Comparative analysis of lengths (in cm) of some LIA knots $\cT_{p,q}$
and unknots given by eqs.  (\ref{sols}).  Total length of the vortex
ring $\cU_0$ of equal diameter is shown for reference (empty diamond).
Values (solid diamond) are plotted versus winding number $w=q/p$.
Torus knots are labeled on the $x$-axis by ($p,q$), whereas for
unknots (bottom diagram) toroidal coils $\cU_{m,1}$ are shown on the
left and poloidal coils $\cU_{1,m}$ are shown on the right, for
$m=1,2,\ldots,9$.}
\label{length}
\end{figure}

As we see from the examples shown in Figure \ref{length}, the total
length monotonically decreases with increasing winding number when
$p>q$, and monotonically increases when $q>p$.  This behaviour is
confirmed for a large class of torus knots tested (not shown in
figure) and seems a generic feature of knot types.  Interestingly,
this behaviour is also confirmed for torus unknots, as evidenced by
the results on the bottom diagram of Figure \ref{length}.

By interchanging $p$ and $q$ the knot type remains the
same ($\cT_{p,q}\sim\cT_{q,p}$); the minimum crossing number 
$c_{\rm min}$ of the knot can be computed by this formula:
\begin{equation}
    c_{\rm min}=(\min{(p,q)}-1)\cdot\max{(p,q)}\ ,
    \label{cmin}
\end{equation}
(for $\cT_{p,q}$ with $1<p<q$, consider minimal knot diagram; then
$c_{\rm min}=(p-1)q$).  The minimum crossing number is the standard
measure of topological complexity and provides also a lower bound on
the average number of apparent crossings $\bar{C}$, a standard
algebraic measure of structural complexity of space curves (Ricca,
2005).  This latter has been proven useful to quantify complex vortex
tangles in space (Barenghi, \emph{et al.}, 2001) and here it may be
used to measure the complexity of torus unknots.  By definition of
minimum crossing number, we evidently have: $c_{\rm min}\le\bar{C}$.
From eq.  (\ref{cmin}) we see that since topological complexity
increases with $p$ and $q$, total length is actually monotonically
increasing with $c_{\rm min}$.  A similar behaviour is found for torus
unknots, where now total length appears to be a monotonic function of
the average crossing number $\bar{C}$.

These results appear to be generic and they seem to be in good 
agreement with current work on properties of ideal knots, where 
possible relationships between $c_{\rm min}$ and minimal length of 
tight knots are envisaged. In the context of LIA knots, our results 
seem to indicate that indeed any increase in complexity, measured by 
an additional crossing $\Delta c_{\rm min}$, is associated with some 
length increase $\Delta L$, that in turn must involve some additional 
energy to the vortex system. In the next section we shall explore 
how kinetic energy actually relates to knot complexity.

\section{Kinetic energy of torus knots and unknots}
A comparative study of the kinetic energy of torus knots and unknots
is done by numerical integration of the Biot-Savart law of solution
equations (\ref{sols}).  Numerical implementation of the Biot-Savart
law is done by standard discretization of the axis curve into $N=300$
segments and standard de-singularization by application of a cut-off
technique (for details, see, for example, Schwarz, 1988; Aarts \&
deWaele, 1994), the only difference here is that dissipation is being
neglected.  This is physically equivalent to consider a quantised 
vortex system in superfluid $^4$He at temperature below $1~\textrm{K}$, 
so that mutual friction (Barenghi \emph{et al.}, 1982) can be neglected, 
and vortex motion in between vortex reconnection events is entirely 
governed by the classical Euler equations (Barenghi, 2008).
The circulation of such vortices is $\Gamma=10^{-3}~\textrm{cm}^2/\textrm{s}$ 
(the ratio of Planck's constant and the helium mass).  For consistency, 
the cut-off used here is based on the superfluid vortex core radius, 
$a \approx 10^{-8}\rm cm$, which is essentially the superfluid coherence 
length. Numerical code is based on the implementation of a source 
code developed by one of us (S. Alamri, \emph{Ph.D. Thesis}, Newcastle U., 
in preparation).

The kinetic energy (per unit density) is given by eq. (\ref{kinetic}).
Note that the density of superfluid helium below $1~\textrm{K}$ 
($=0.145~\textrm{g}/\textrm{cm}^4$) is constant, so hereafter we shall refer 
to the "kinetic energy per unit density" simply as "kinetic energy". 
The numerical computation of the volume integral of eq.
(\ref{kinetic}) is not practical; it is more convenient to reduce the
volume integral to a line integral (for its derivation see, for
example, Barenghi \emph{et al.}, 2001):
\begin{equation}
T=\frac{\Gamma}{2}\oint_{\cC}\bu\cdot\bX\times\bt\, \rd s\ ,
\label{kineticline}
\end{equation}
where, as before, vorticity contribution is expressed through the 
vortex circulation $\Gamma$.

\begin{figure}[t!] 
\centering     
\includegraphics[width=0.7\textwidth]{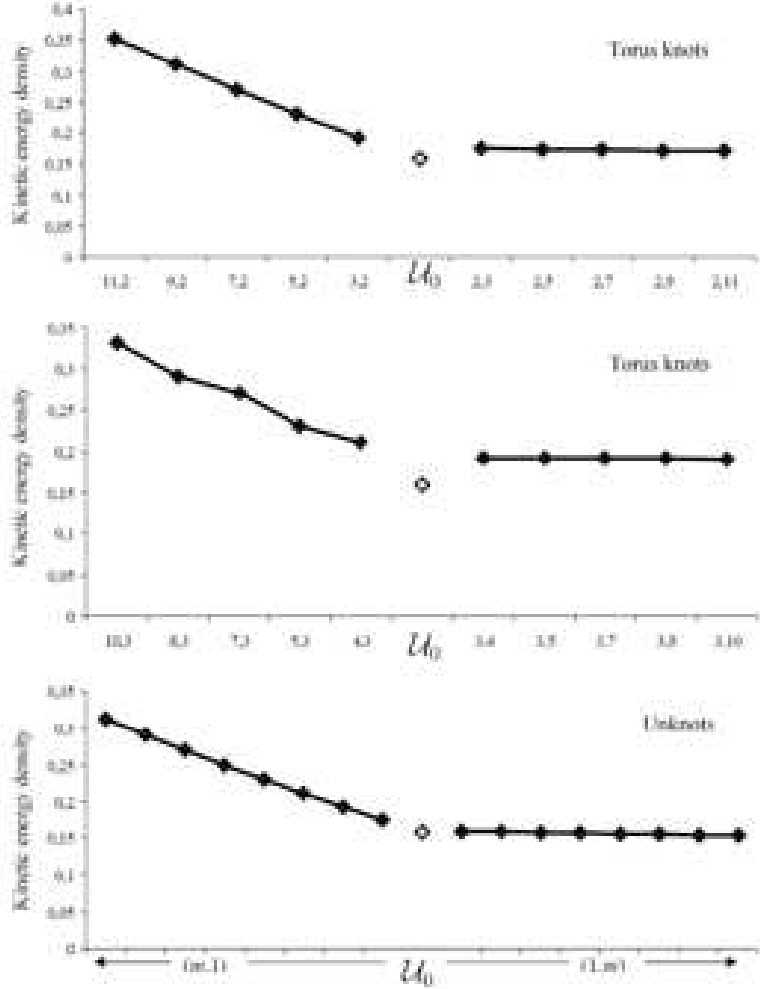}
\caption{Comparative analysis of kinetic energy density of some LIA
knots $\cT_{p,q}$ and unknots given by eqs.  (\ref{sols}).  Kinetic
energy density of the vortex ring $\cU_0$ of equal diameter is shown
for reference (empty diamond).  The kinetic energy, given by eq.
(\ref{kinetic}), has dimensions $\textrm{cm}^5/\textrm{s}^2$, hence
$T/L$ has dimensions $\textrm{cm}^4/\textrm{s}^2$.  Values (solid
diamond) are plotted versus winding number $w=q/p$.  Torus knots are
labeled on the $x$-axis by ($p,q$), whereas for unknots (bottom
diagram) toroidal coils $\cU_{m,1}$ are shown on the left and poloidal
coils $\cU_{1,m}$ are shown on the right, for $m=1,2,\ldots,9$.}
\label{energylength}
\end{figure}

Since kinetic energy is constant during evolution, we need
not compute it as a function of time. Data on kinetic energy are 
collected for several torus knots, given by
$\{p,q\}=2,3,4,\ldots,11$, and unknots, given by $m=1,2,3,\ldots,9$.
Kinetic energy of the vortex ring of same diameter $\cU_{0}$ is
reported for comparison.

Let us first consider the kinetic energy per unit length (kinetic
energy ``density'').  This is given by dividing the kinetic energy of
each knot configuration by the corresponding knot length.  We present
results for the two families of knots previously considered and 
unknots, but similar results have been found for other families of 
knots not presented here.  The dependence of kinetic energy density on
winding number shown for these two knot families (see top diagrams of 
Figure \ref{energylength}) is qualitatively similar to that found for the
other families of torus knots tested.  Results show a
marked difference between right- and left-hand-side data
distributions.  Kinetic energy density of knot configurations given by
$w>1$ seem to be constant or slightly decreasing with $w$. Remember 
that this vortex knots are LIA stable.  Knot configurations given by 
$w<1$ show, on the contrary, that kinetic energy density increases 
with increasing knot complexity. Similar behaviour is found for unknots
(bottom diagram of Figure \ref{energylength}): kinetic 
energy density of poloidal coils, that are LIA stable, seem to be 
almost constant (or even slightly decreasing with the winding number), 
whereas toroidal coils show a marked monotonic dependence on 
increasing winding number. 

A possible justification of this is to interpret kinetic energy by using 
the LIA law; by using eq. (\ref{lia}), we have
\begin{equation}
    T_{\rm LIA}=\frac{1}{2}\int_{V}\|\bu_{\rm LIA}\|^2\,\rd\bx^{3}
               \propto\oint_{\cC}c^{2}\,\rd s=cst.\ ,
    \label{kineticlia}    
\end{equation}
where the latter equation is due to a geometric interpretation of the
conservation laws associated with LIA (Ricca, 1993).  Since under LIA
$L$ is also constant for each vortex filament, from the
right-hand-side of eq.  (\ref{kineticlia}) we see that the energy
density is, to first approximation, constant for each vortex and
proportional to $c^{2}$.  For small-amplitude torus knots (and
unknots) we can estimate how curvature varies with $w$; for simplicity
consider first the case of the unknots: for toroidal coils ($w<1$) the
radius of curvature $R$ evidently decreases with increasing $m$;
since, to first approximation, this is also true for torus knots (by
fixing $q$ and replacing $m$ by $p$), we have that $c=R^{-1}$ tends to
decrease with increasing $w$.  For poloidal coils ($w>1$), however,
since $L$ does not vary much with $m$ (see Figure \ref{length}), the
radius of curvature $R$ consequently won't vary much with $m$. This 
is also approximately true for torus knots (by fixing $p$ and replacing 
$m$ by $q$), meaning that $c=R^{-1}$ won't vary much with $w$. Since
kinetic energy density is normalized by knot length, which increases
with complexity, we have the two distinct behaviours observed.

\begin{figure}[t!] 
\centering     
\includegraphics[width=0.7\textwidth]{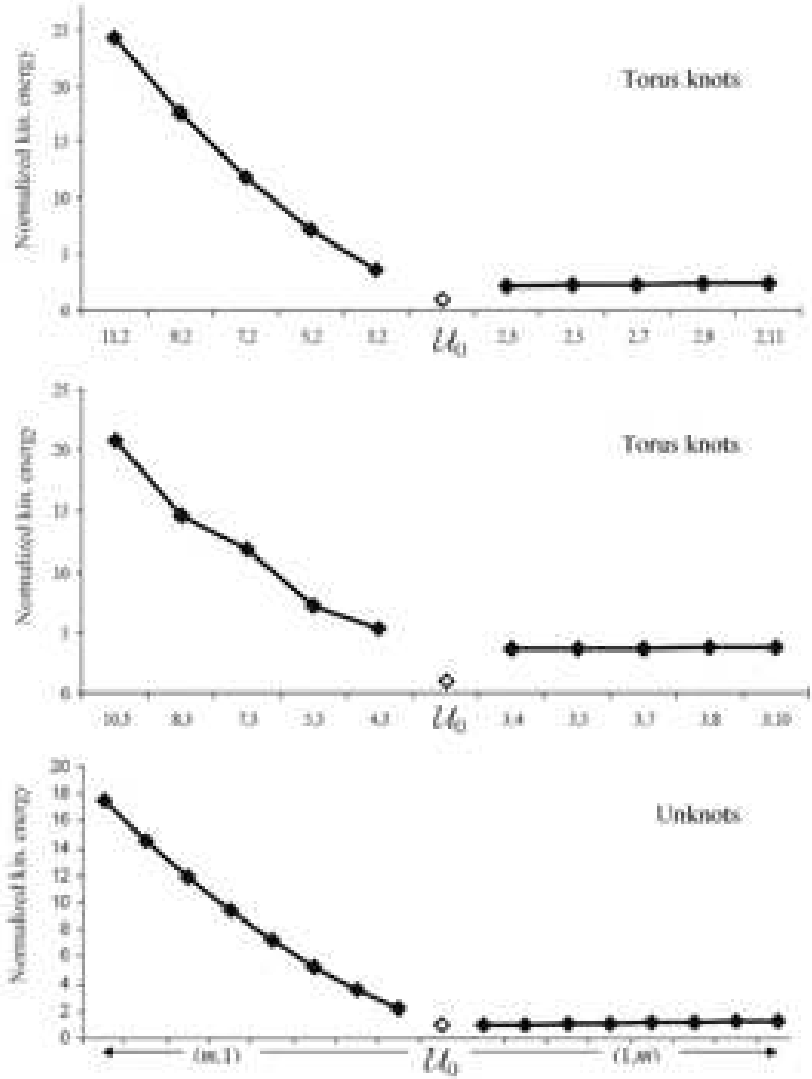}
\caption{Comparative analysis of normalized kinetic energy of some LIA
knots $\cT_{p,q}$ and unknots given by eqs.  (\ref{sols}).  Kinetic
energy is referred to that of the vortex ring $\cU_0$ of same diameter
(empty diamond).  Values (solid diamonds) are plotted versus winding
number $w=q/p$.  Torus knots are labeled on the $x$-axis by ($p,q$),
whereas for unknots (bottom diagram) toroidal coils $\cU_{m,1}$ are
shown on the left and poloidal coils $\cU_{1,m}$ are shown on the
right, for $m=1,2,\ldots,9$.}
\label{energynorm}
\end{figure}

Finally, let us consider the kinetic energy simply normalized by the 
reference energy of the vortex ring of same diameter. Results are 
shown in Figure \ref{energynorm}. As we can see the normalize total 
energy increases with knot complexity, markedly for $w<1$, but 
also very slightly for $w>1$. Diagrams clearly indicate that more 
complex knots have higher energy than simpler one, the vortex ring 
having the lowest kinetic energy. 

\section{Discussion of results}
Figure \ref{energylength} and \ref{energynorm} show that the kinetic 
energy per unit length can change by more than 100 percent for both 
knots and unknots, depending on the particular vortex configuration. 
This result has implications for the study of quantum turbulence at 
very low temperatures in superfluid $^4$He, a problem which is currently 
receiving experimental (Walmsley \emph{et al.}, 2007)
and theoretical attention (Alamri \emph{et al.}, 2008).

It is known that quantum turbulence consists of a
disordered tangle of vortex filaments, and that, even at high
temperatures in the presence of friction which damps out
kinks and Kelvin waves along the filaments, the geometry of
the tangle is fractal (Kivotides \emph{et al.}, 2001).
There is current interest in measuring the energy of quantum
turbulence and its temporal decay at very low temperatures, because
it appears that, without the friction, a Kelvin--wave cascade process
(Kivotides \emph{et al.}, 2001), analogous to the classical Kolmogorov 
cascade, can shift the kinetic energy to such high wave-numbers that 
kinetic energy can be radiated away acoustically. Unfortunately the main 
experimental techniques available (based on second--sound and ion--trapping) 
actually measure the length of the vortex filaments, not the energy.
This is why the relation between the length of a vortex
(which can be detected directly) and its kinetic energy 
(which cannot) is important. The natural question to ask is whether
Kelvin waves of shorter and shorter wavelength can be added to a
vortex filament without altering its energy, as the velocity fields
of neighboring vortex strands cancel each other out. Our study
of vortex knots makes a step forwards in answering this question.

The kinetic energy per unit length of a {\it straight} vortex filament 
is easily computed:
\begin{equation}
   T/L=\frac{1}{2} \int_{V}\|\bu\|^2\, \rd \bx^3
      =\frac{\Gamma^2}{4 \pi^2} \ln\frac{b}{a} \ ,
\end{equation}
where we used cylindrical polar coordinates and the fact that
$\|\bu\|=\Gamma/(2 \pi r)$; here the upper cut-off $b$ represents 
either the radius of the container or the distance to the nearest vortex.
This expression is often used in the helium literature to estimate that 
the energy density (energy per unit volume) of a turbulent vortex tangle 
of measured vortex line density $\Lambda$ (vortex length per unit volume) is
$(\Gamma^2 \Lambda)/(4 \pi^2) \ln{b/a}$ where $b=\Lambda^{-1/2}$.
Figures \ref{energylength} and \ref{energynorm} show that the kinetic energy 
per unit length is not the same, even for relatively simple vortex 
configurations, hence the above estimate for the energy density of the
turbulent flow must be used with care.

\section*{Acknowledgemts}
\label{sec:Acknowledgemts}

F. Maggioni acknowledges financial support from the grant ''Fondi di 
Ricerca di Ateneo di Bergamo 2008: Rilassamento magnetico mediante 
meccanismi STF''.


\end{document}